# Gravitational Instabilities In Disks With Radiative Cooling

Annie C. Mejía, Richard H. Durisen
*Astronomy Department. Indiana University. 727 E. 3rd St. Bloomington, IN 47401*

Megan K. Pickett
*Department of Chemistry and Physics, Purdue University Calumet. 2200 169th Street, Hammond, IN 46323*

**Abstract.** Previous simulations of self-gravitating protostellar disks have shown that, once developed, gravitational instabilities are enhanced by cooling the disk constantly during its evolution (Pickett et al. 2002). These earlier calculations included a very simple form of volumetric cooling, with a constant cooling time throughout the disk, which acted against the stabilizing effects of shock heating. The present work incorporates more realistic treatments of energy transport. The initial disk model extends from 2.3 to 40 AU, has a mass of 0.07 $M_\odot$ and orbits a 0.5 $M_\odot$ star. The models evolve for a period of over 2500 years, during which extensive spiral arms form. The disks structure is profoundly altered, transient clumps form in one case, but no permanent bound companion objects develop.

## 1. Introduction

The formation of long-lived clumps by gravitational instabilities in simulations of protoplanetary disks, and therefore the survival of possible planets formed in this fashion, depends critically on the treatment of the disk's thermal physics, specifically EOS, heating, cooling, and energy transport (Pickett et al. 2002; Nelson, Benz, & Ruzmaikina 2000; Boss 2002). We present three calculations with different algorithms for vertical radiative cooling which were carried out in order to study their effect on the thermal structure of the disk, the development of the instabilities, and the formation of clumps or possible protoplanets.

## 2. Simulations

The initial model for all three simulations is similar to that described in Durisen et al. (2001), but scaled to have a central star of 0.5 $M_\odot$ and a nearly Keplerian disk of 0.07 $M_\odot$, with surface density $\Sigma(r) \propto r^{-0.5}$ from 2.3 to 40 AU (Figure 1). Our second-order 3D code (Pickett et al. 1998) solves the hydrodynamics equations in conservative form on a cylindrical grid of uniform spacing with r,$\phi$,z-resolution of 512,128,32 and reflection symmetry about z = 0. The initial equilibrium model extends from radial zone number 16 (2.3 AU) to 242 (40 AU), the extra radial cells accommodate expansion of the disk due to the instabilities. All three calculations include the effects of shock heating through the use of an artificial bulk viscosity, but have different forms of vertical radiative cooling as described below. Molecular weights, Rosseland mean opacities, and

Plank mean opacities as functions of temperature and pressure for Cases 2 and 3 were provided by P. D'Alessio & N. Calvet.

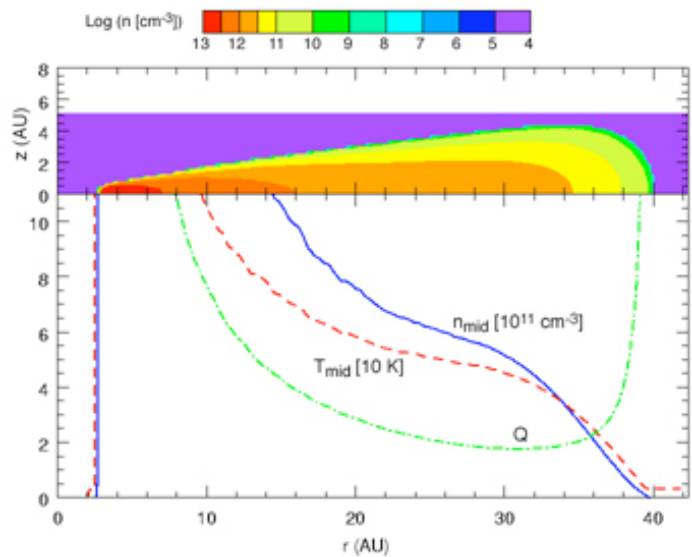

Figure 1. Initial model. The top panel shows meridional contours of number density. The zero-density surface is almost identical to the $10^8$ cm$^{-3}$ contour shown. The bottom panel shows the radial distribution of the midplane number density $n_{mid}$ in units of $10^{11}$ cm$^{-3}$, the midplane temperature $T_{mid}$ in units of 10 K, and the Toomre Q.

Case 1: Constant cooling time. The volumetric cooling rate is set each time step so that the midplane cooling time $t_{cool}$ is 2 ORPs (1 ORP = the typical outer rotation period of the initial disk = 250 years at 33 AU).

Case 2: Eddington atmosphere. The volumetric cooling rate is set so that a vertical column in the disk radiates with an effective temperature computed using an Eddington grey atmosphere fitted to the column's midplane temperature and normal optical depth. The cooling times from this approach are ≥ 1 ORP over most of the disk's midplane.

Case 3: Flux limited diffusion plus Eddington atmosphere. Fluxes are calculated in the optically thick part of a column using the radiative diffusion algorithm from Bodenheimer et al. (1990). An atmosphere is fitted to the optically thin part of the disk, where the effective temperature of the column is calculated in a similar fashion to Case 2, but using the temperature of the topmost optically thick cell instead of the midplane temperature. Initial cooling times in the midplane are ≤ 1 ORP.

## 3. Results

All three cases evolve for at least 2500 yr, developing spiral instabilities between 1150 and 1500 yr. After the onset of the instabilities, the highest relative heating rate due to shocks in a column occurs near the surface of the disks. Therefore, in general, the gas at the surface has a higher temperature than at the midplane in the same column. Radially and azimuthally, most of the heating and cooling occurs in the regions leading the spiral arms, where the gas gets compressed, and in the outer disk, where low density gas falls back radially after the strong spiral arm phase.

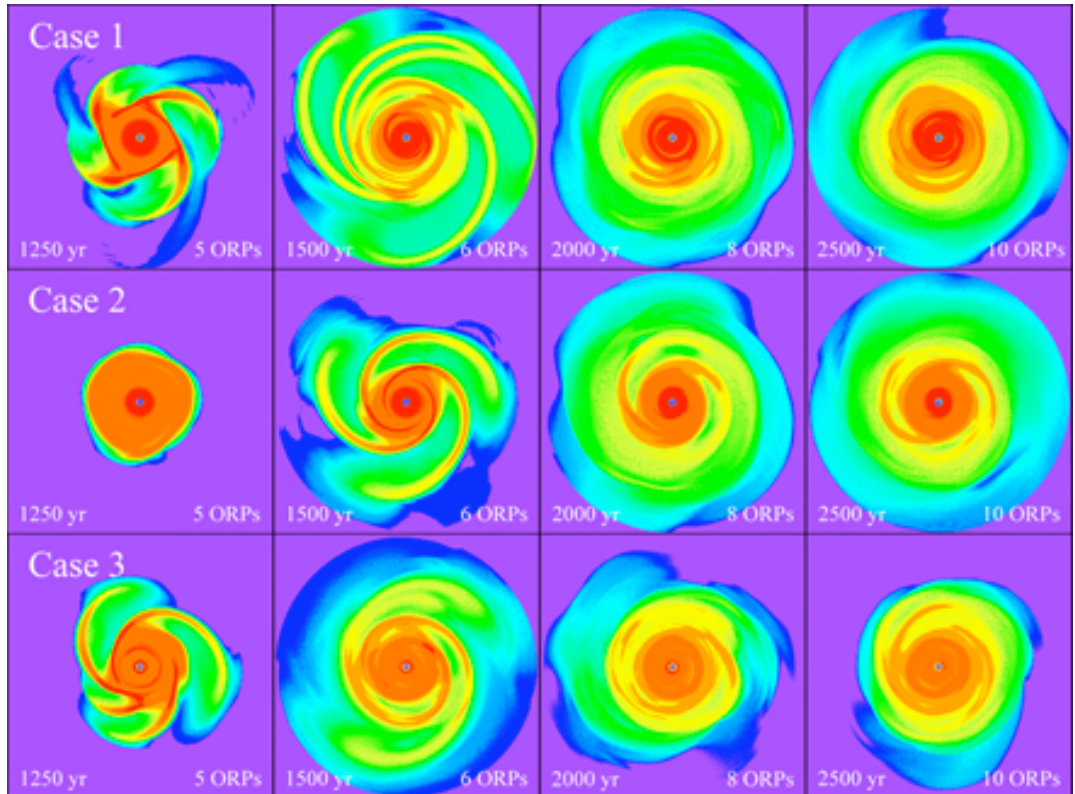

Figure 2.  Sequence of snapshots of midplane number densities for all the cases. The color scale is the same as in Figure 1.  Each panel is 170 AU on a side.

Case 1:  After 1800 yr of evolution, this disk develops a system of dense concentric rings and elongated clumps at 7, 10, and 13 AU (Figures 2 and 3) which survive for at least another 2300 yr.  The features become more concentrated as the disk evolves.  Dense clumps do form in the rings but seldom last for more than an orbit.

Case 2:  After 2750 yr of evolution, typical cooling times are about 8 ORPs at the disk's midplane.  The spiral structure is not as strong as in Case 1 and is mostly in the form of two loosely wound spiral arms.  No long-surviving clumps develop.

Case 3:  The typical cooling times at the disk's midplane are about 10 ORPs after 2500 yr.  A high mass accretion event happens between 1500 and 1800 yr when $\sim 5.5 \times 10^{-3}$ $M_\odot$ crosses the inner boundary toward the center of the grid.  This corresponds to inward mass transport at a rate of about $2 \times 10^{-6}$ $M_\odot$/yr.  The accretion causes the lack of structure in the inner disk (Figure 2).

## 4.  Discussion

Even with the same initial conditions, the three cases in this study have shown that the outcome of their evolution can be quite different depending on the details of how they are cooled.  Despite being the least physically realistic of the cooling schemes, Case 1 demonstrates that efficient cooling along the columns and throughout the entire evolution of the disk is needed to produce transient dense clumps.  Whether these clumps will eventually survive long enough to coalesce into planets is not yet known, but see Mayer et al. (2003, these proceedings) for a simulation relevant to this issue.  Azimuthal mass averages show that, even though the clumps rarely last longer than an orbital period, newly formed ones always follow the same stable, circular orbits.

These orbits correspond to the growing dense mass rings at 7, 10, and 13 AU. Case 2 was meant to be a transition between Cases 1 and 3 and was the first of our simulations to use opacities and proper molecular weights. A flaw in our algorithm for Case 2 produced an unwanted source of heating in low optical depth regions, and therefore we are cautious about making strong claims from this case. We consider the handling of Case 3 at high optical depths to be better, but the simple Eddington atmosphere fit above the disk photosphere does not properly cool the disk's upper layers, where much of the shock heating occurs. Very recently we have developed another way of treating the cooling at low optical depths in upper layers heated by both shocks and stellar irradiation. Testing of new routines for optically thin regions and for stellar irradiation are under way.

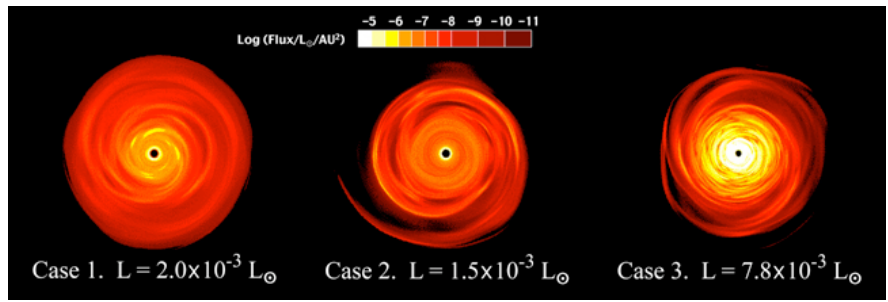

Figure 3. Flux maps in $L_\odot/AU^2$ after 2500 yr of evolution. The total luminosities are low compared to those of typical T Tauri disks because the hot central 2.3 AU region is missing. Notice the clumps at 13 AU in Case 1 and the luminous inner disk in Case 3 due to heating associated with the accretion event. Each panel is 170 AU on a side.